\documentclass[sigconf]{acmart}
\acmConference[WSDM '27]{The 20th ACM International Conference on Web Search and Data Mining}{February 2027}{}
\AtBeginDocument{%
  }

\usepackage{booktabs}
\usepackage{graphicx}
\usepackage{amsmath}
\usepackage{multirow}

\begin{document}

\title{ICEGR: An Intent-Coherent End-to-End Generative Retrieval Framework for E-commerce Search}


\settopmatter{authorsperrow=3}

\author{Jiayi Tuo}
\authornote{These authors contributed equally to this research.}
\email{tuojiayi@baidu.com}
\affiliation[obeypunctuation=true]{%
  \institution{Baidu},
  \city{Beijing},
  \country{China}
}

\author{Hehan Li}
\authornotemark[1]
\email{hehanli@baidu.com}
\affiliation[obeypunctuation=true]{%
  \institution{Baidu},
  \city{Beijing},
  \country{China}
}

\author{Dongjun Fu}
\authornotemark[1]
\email{fudongjun@baidu.com}
\affiliation[obeypunctuation=true]{%
  \institution{Baidu},
  \city{Beijing},
  \country{China}
}

\author{Xin Lu}
\email{luxin06@baidu.com}
\affiliation[obeypunctuation=true]{%
  \institution{Baidu},
  \city{Beijing},
  \country{China}
}

\author{Ling Zhuang}
\email{zhuangling@baidu.com}
\affiliation[obeypunctuation=true]{%
  \institution{Baidu},
  \city{Beijing},
  \country{China}
}

\author{Fuwei Zhang}
\email{zhangfuwei@buaa.edu.cn}
\affiliation[obeypunctuation=true]{%
  \institution{Beihang University},
  \city{Beijing},
  \country{China}
}

\author{Meifang Li}
\email{limeifang@baidu.com}
\affiliation[obeypunctuation=true]{%
  \institution{Baidu},
  \city{Beijing},
  \country{China}
}

\author{Peizhi Xu}
\email{xupeizhi@baidu.com}
\affiliation[obeypunctuation=true]{%
  \institution{Baidu},
  \city{Beijing},
  \country{China}
}

\author{Hanmeng Liu}
\email{liuhanmeng@baidu.com}
\affiliation[obeypunctuation=true]{%
  \institution{Baidu},
  \city{Beijing},
  \country{China}
}

\author{Shuanglong Li}
\email{lishuanglong@baidu.com}
\affiliation[obeypunctuation=true]{%
  \institution{Baidu},
  \city{Beijing},
  \country{China}
}

\author{Liwei Qian}
\email{qianliwei@baidu.com}
\affiliation[obeypunctuation=true]{%
  \institution{Baidu},
  \city{Beijing},
  \country{China}
}

\author{Yanbiao Ma}
\correspondingauthor
\email{ybma1998@ruc.edu.cn}
\affiliation[obeypunctuation=true]{%
  \institution{Renmin University of China},
  \city{Beijing},
  \country{China}
}

\author{Fuzhen Zhuang}
\correspondingauthor
\email{zhuangfuzhen@buaa.edu.cn}
\affiliation[obeypunctuation=true]{%
  \institution{Beihang University},
  \city{Beijing},
  \country{China}
}

\renewcommand{\shortauthors}{Tuo et al.}

\begin{abstract}
Generative Retrieval (GR) is promising for e-commerce search, yet existing methods struggle to maintain query-intent consistency throughout the training pipeline. First, semantic ID (SID) construction based on static product information limits SIDs' ability to encode product--intent associations. Second, although supervised fine-tuning (SFT) learns product--SID mappings across the catalog, low-exposure products still lack real query-intent supervision because query-to-SID training relies solely on online logs, leading to poor retrieval performance for these products. Third, business-oriented preference optimization may favor popular or high-value products over those best matching the query intent, weakening query--product relevance.
To address these issues, we propose \textbf{ICEGR}, an Intent-Coherent End-to-End Generative Retrieval Framework for E-commerce Search, which integrates query intent throughout GR training. ICEGR comprises three components: (1) \textbf{Intent-Aware SID Construction}, which incorporates query-intent signals into SID construction, enabling SIDs to capture search intent beyond static product information; (2) \textbf{Synthetic Query-Enhanced Unified SFT}, which unifies multiple SFT tasks under a single query-to-SID objective and augments sparse supervision from online logs with synthetic queries, providing complementary query-intent supervision for low-exposure products; and (3) \textbf{Relevance-Calibrated Preference Optimization}, which integrates
query--product relevance and business signals into a margin-adaptive
preference objective, preserving query intent while enabling business
preference learning. Offline results show that ICEGR improves \textbf{Recall@$20$} by \textbf{21.7\%} and \textbf{NDCG@20} by \textbf{26.6\%}, respectively, over the baseline. When deployed as an \textbf{end-to-end} generative retrieval pathway in Baidu E-commerce Search system, ICEGR achieves relative improvements of \textbf{3.52\%} in \textbf{CTR}, \textbf{15.96\%} in \textbf{order volume}, and \textbf{7.53\%} in \textbf{GMV} in an online A/B test.
\end{abstract}



\maketitle

\begin{figure}[htbp]
  \centering
  \includegraphics[
    width=\columnwidth
  ]{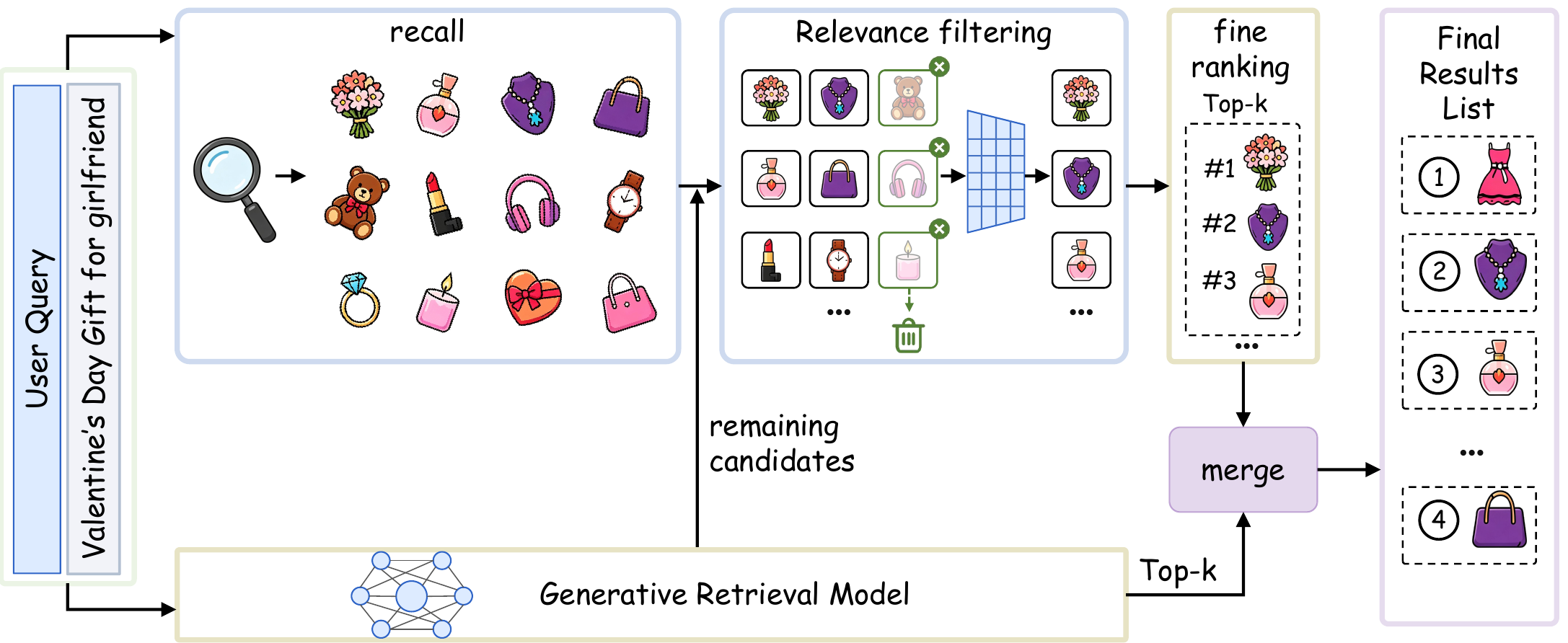}
  \caption{Traditional MCA versus generative retrieval. Traditional MCA
  recalls, filters, and ranks product candidates, whereas ICEGR directly
  generates the top-$k$ products. The two candidate sets are then merged to
  form the final results list.}
  \Description{Comparison of traditional MCA and generative retrieval.
  Traditional MCA recalls, filters, and ranks product candidates. ICEGR
  directly generates top-k products. The two candidate sets are merged into
  the final results list.}
  \label{fig:traditional-mca-generative-retrieval}
\end{figure}


\section{Introduction}
\label{sec:introduction}

E-commerce search aims to retrieve, from a large-scale catalog, products that
satisfy the shopping intent expressed in a user query~\cite{10.1145/3447548.3467101}. As illustrated in Figure~\ref{fig:traditional-mca-generative-retrieval}, industrial e-commerce search systems commonly adopt a multi-stage cascade architecture (MCA), consisting of query understanding, retrieval, pre-ranking, and ranking stages~\cite{10.1145/3097983.3098011,10.1145/3770855.3818456}. By progressively narrowing the candidate set, this architecture produces the final ranked product list. However, different stages often rely on separate models, features, and optimization objectives. Once a relevant product is missed during retrieval, downstream ranking models have limited ability to recover it, thereby placing an upper bound on the overall system performance~\cite{10.1145/3477495.3532050}.

Generative retrieval provides a promising direction for addressing the limitations of cascade architectures. This paradigm maps each product to a generative semantic ID (SID) and employs an autoregressive model to directly generate the SIDs of relevant products conditioned on a user query, thereby mitigating the performance loss caused by cascading multiple stages~\cite{3600270.3601857,10.1145/3774904.3792862}. However, although existing end-to-end models unify the retrieval architecture, they struggle to maintain query-intent consistency throughout the learning pipeline~\cite{10.1145/3726302.3731951,10.1145/3774904.3792562}. Across different stages, these models may favor product content, product-oriented supervision signals, or popularity signals, rather than consistently modeling the actual query--product relevance relationship~\cite{yang-etal-2025-gsid,10.1145/3442381.3450129,10.1145/3770855.3818495}. This gap between architectural unification and query-intent consistency remains a key obstacle to the effective deployment of generative retrieval in industrial systems.

More specifically, insufficient query-intent consistency manifests in three
aspects. First, during SID construction, existing methods typically construct
SIDs from static product content, such as titles and descriptions~\cite{yang-etal-2025-gsid,10.1145/3774904.3792862}. However,
content similarity does not fully capture how products relate to query intent~\cite{10.1145/3701716.3715228}.
For example, the query ``Valentine's Day gift for girlfriend'' may be satisfied
by either a Chanel lipstick or a Bulgari necklace. Although these products
differ substantially in content, they satisfy the same shopping intent. Second, during supervised learning, product-centric supervision primarily
encourages the model to memorize the mapping between products and their SIDs
rather than learn to retrieve relevant SIDs from query intent~\cite{3600270.3601857,10.1145/3774904.3792862}. Although
real-world interaction logs provide query-to-SID supervision~\cite{10.1145/3805712.3808479}, product exposure
in industrial e-commerce systems follows a long-tailed distribution.
Consequently, many low-exposure products lack sufficient and reliable
query--product association signals~\cite{10.1145/3805712.3808409}. Finally, during preference optimization,
directly constructing preference pairs from click-through rates or business
value may bias the model toward popular or high-value products that do not match
the query intent, further exacerbating intent misalignment~\cite{10.1145/3805712.3808512,10.1145/3770855.3818495}.

To address these issues, we propose \textbf{ICEGR}, an Intent-Coherent End-to-End
Generative Retrieval Framework for E-commerce Search. ICEGR places query intent
at the center of the entire generative retrieval learning pipeline. First, we
introduce \textbf{Intent-Aware SID Construction (IA-SID)}, which incorporates
query-induced product co-click relationships and historically associated
queries before product discretization. This enables product SIDs to encode both
content semantics and retrieval-intent associations. Second, we propose
\textbf{Synthetic Query-Enhanced Unified SFT (SQE-SFT)}, which
automatically constructs multi-granularity synthetic queries from the entire
product catalog and replaces product-memorization training with synthetic
query-to-SID supervision. It reduces the mismatch between training and
inference inputs and supplements interaction-log supervision for low-exposure
products, improving their retrievability. Finally, we propose
\textbf{Relevance-Calibrated Preference Optimization (RCPO)}, which integrates query--product relevance and business signals
into preference optimization. RCPO restricts preference comparisons to
semantically relevant candidates, refines their ordering using
user-behavior and business-value signals, and adapts each preference
update to the pairwise composite-score margin. This design preserves
query intent while enabling business preference learning.

We evaluate ICEGR offline and in a production-scale online A/B test. Offline,
ICEGR improves Recall@$20$ and NDCG@$20$ by 21.7\% and 26.6\%, respectively. We further deploy ICEGR as an end-to-end
generative retrieval pathway in Baidu E-commerce Search system. The online
A/B test yields relative gains of 3.52\% in CTR, 15.96\% in order volume, and
7.53\% in GMV. These results demonstrate that maintaining query-intent
consistency across SID construction, generative supervision, and preference
optimization can translate offline retrieval improvements into tangible online
business gains. More broadly, this intent-centric, pipeline-wide optimization
paradigm moves beyond stage-wise local optimization and provides a practical
approach to jointly improving retrieval quality and business outcomes.

In summary, our main contributions are as follows:

\begin{enumerate}

  \item We propose Intent-Aware SID Construction, which integrates
query-induced product co-click relationships and historically associated
queries before product discretization. SIDs encode product semantics
while preserving query--product intent associations revealed by search
behavior.

  \item We introduce Synthetic Query-Enhanced Unified SFT, which automatically
  constructs multi-granularity synthetic queries across the catalog and
  reformulates conventional product-memorization training as query-to-SID
  supervision. This provides low-exposure products with complementary
  query-level supervision beyond sparse interaction logs.

  \item We propose Relevance-Calibrated Preference Optimization, which
integrates semantic relevance constraints, behavioral signals, and
business value into preference learning. By calibrating preference
updates with pairwise composite-score margins, RCPO preserves query
intent while enabling business preference learning.

  \item We deploy ICEGR in Baidu E-commerce Search system as an end-to-end
  generative retrieval pathway. Offline experiments and a
  production-scale A/B test validate its effectiveness in improving
  retrieval quality and business outcomes.
\end{enumerate}

\begin{figure*}[t]
    \centering
    \includegraphics[width=\textwidth]{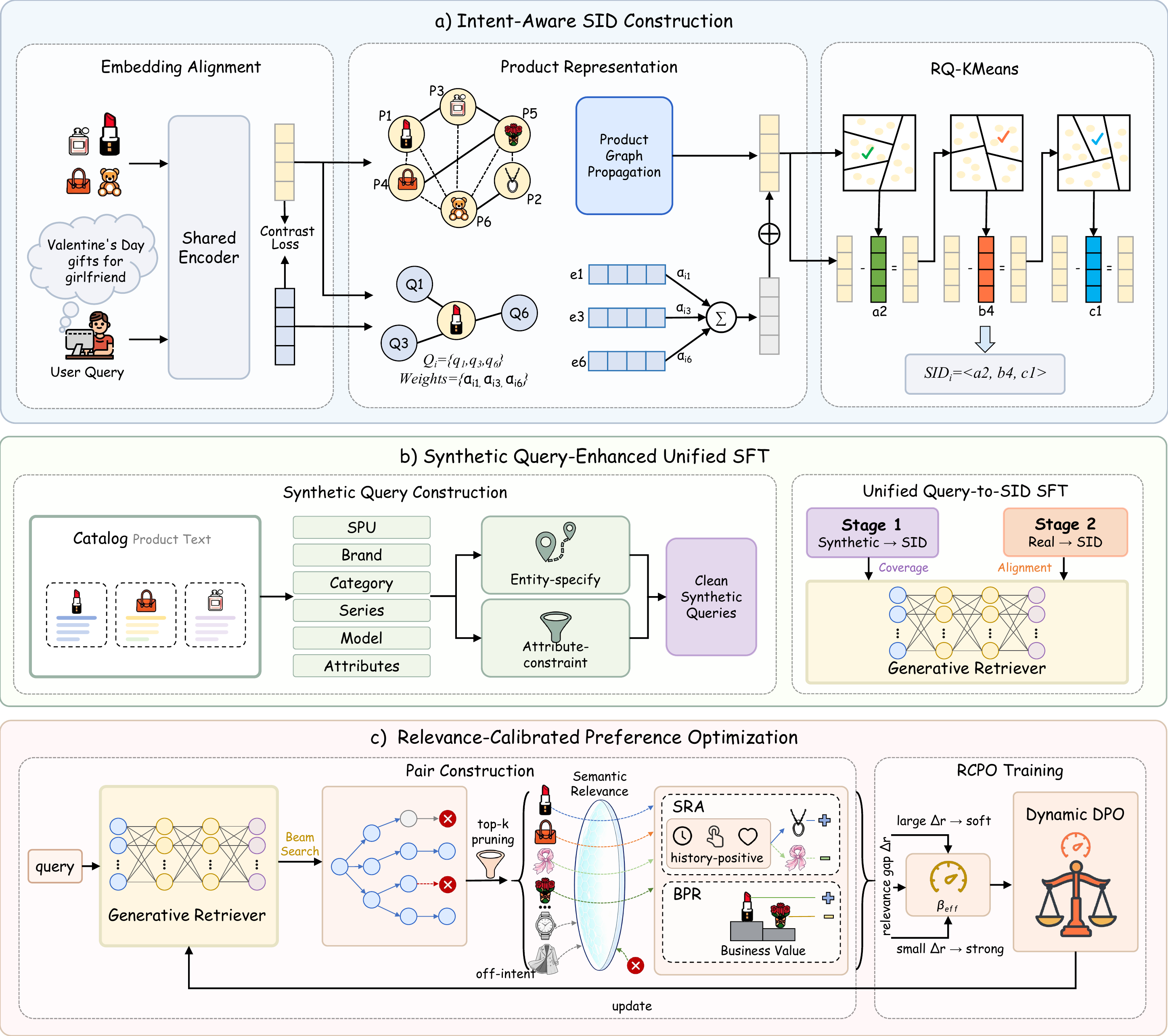}
    \caption{Overview of ICEGR, which enforces query-intent alignment throughout generative retrieval in three stages: (a) IA-SID constructs semantic IDs aligned with search intent; (b) SQE-SFT expands sparse real query-to-SID supervision with multi-granularity synthetic queries, followed by fine-tuning on real queries under the same objective; and (c) RCPO learns business-aware preferences among semantically relevant candidates through margin-adaptive preference optimization.}
    \Description{Overview of the three-stage ICEGR framework.
IA-SID aligns product semantic IDs with search intent using
query-product interaction signals. SQE-SFT constructs
multi-granularity synthetic queries from catalog semantics, trains on
synthetic query-to-SID pairs, and then fine-tunes on real query-to-SID
interactions using the same objective. RCPO combines semantic
relevance with behavioral and business signals to learn business-aware
preferences while preserving query-intent alignment.}
    \label{fig:icegr-framework}
\end{figure*}

\section{Related Work}

\subsection{E-commerce Search and Product Retrieval}

E-commerce search retrieves products from large-scale catalogs according to the shopping intent expressed in user queries~\cite{10.1145/3447548.3467101}. Industrial systems typically adopt multi-stage cascades that combine query understanding, candidate retrieval, pre-ranking, and fine ranking to balance retrieval efficiency and ranking quality~\cite{10.1145/3097983.3098011,10.1145/3770855.3818456,10.1145/3805712.3808470}. These stages are often optimized with different models and objectives~\cite{10.1145/3477495.3532050}. Embedding-based models such as DSSM have been widely used for candidate retrieval~\cite{10.1145/2505515.2505665}, while ranking models such as DeepFM, DCN, and DIN have been developed for feature-interaction modeling~\cite{3172077.3172127,10.1145/3124749.3124754,10.1145/3219819.3219823}. Although multi-stage cascades are effective in practice, products missed during retrieval cannot be recovered by downstream ranking, and the use of separate objectives can make query--product relevance difficult to maintain consistently across stages.

\subsection{Generative Retrieval for E-commerce Search}

Generative retrieval formulates candidate retrieval as conditional generation, where products or documents are represented by discrete identifier sequences and generated from a query or user context~\cite{3600270.3602126,3600270.3601857,3666122.3666574}. In e-commerce search, generated products must satisfy query--product relevance and product-attribute constraints while accommodating practical business objectives~\cite{10.1145/3701716.3715228,10.1145/3726302.3731951,10.1145/3770855.3818495,10.1145/3774904.3792862}.

Recent studies have investigated generative retrieval from both system
and modeling perspectives. LLMGR applies LLM-based generative retrieval
to industrial search and incorporates external knowledge to improve
retrieval in Alipay Search~\cite{10.1145/3626772.3661364}. OneSearch
develops a unified generative search framework, while OneSearch-V2 investigates latent reasoning and self-distillation
~\cite{chen2026onesearch,
chen2026onesearchv2latentreasoningenhanced}. OneRetrieval consolidates
multiple e-commerce retrieval branches into an editable generative
model~\cite{zhang2026oneretrievalunifyingmultibranchecommerce}. Another line of work focuses on constructing informative semantic
identifiers. GSID learns generative semantic representations for
products, while CAT-ID2 and HierGR incorporate category or hierarchical
structures into identifier learning
~\cite{yang-etal-2025-gsid,10.1145/3773966.3777925,zhang-etal-2025-hiergr}.
CQ-SID further investigates constrained semantic identifier learning
with expert-guided reinforcement learning
~\cite{zhu2026efficientgenerativeretrievalecommerce}. Query and intent
information have been incorporated into generative retrieval. TSGR
studies query-aware semantic representations for Taobao search
~\cite{zhan2026tsgrtaobaosearchgenerative}, while CaLIR investigates
category-guided latent-intent reasoning for generative retrieval
~\cite{zhang2026matchingcategoryguidedlatentintent}. Together, these
methods enrich the identifier space or the generation process with product, query, or intent information. This line of work
motivates a question of how such relevance signals should be
preserved during subsequent training and optimization.

Query-level supervision, relevance learning, and preference optimization
provide complementary ways to improve generative retrieval.
Synthetic-query methods can augment query--product supervision,
particularly for products with limited interaction data
~\cite{10.1145/3805712.3808409}. Reinforcement learning has also been
used to improve query--product relevance modeling in e-commerce search
~\cite{10.1145/3774904.3792799}. Preference optimization methods use
behavioral signals, such as clicks, to align generated results with user
preferences~\cite{li2024generativeretrievalpreferenceoptimization}.
Taken together, prior work has advanced generative retrieval through
system unification, semantic identifier construction, query-level
supervision, and preference optimization. These components play
different roles in the retrieval pipeline: the SID space defines the
generation targets, supervised training learns the query-to-SID mapping,
and preference optimization determines which generated products are
favored. Without a shared relevance criterion, later stages may weaken
the query-intent signal established earlier in the pipeline.

ICEGR addresses this pipeline-level consistency problem by using
query--product relevance as a shared criterion across SID construction,
supervised training, and preference optimization. It incorporates
query-induced product relations and historical query semantics into SID
construction, augments query-to-SID supervision with multi-granularity
synthetic queries, and constrains business preference optimization with
relevance-calibrated comparisons.

\section{Method}
This section presents the ICEGR framework, as illustrated in Fig.~\ref{fig:icegr-framework}. 
The framework comprises three sequential stages: IA-SID (Sec.~\ref{sec:ia-sid}), 
SQE-SFT (Sec.~\ref{sec:sqe-sft}), and RCPO (Sec.~\ref{sec:rcpo}).

\subsection{Intent-Aware SID Construction}
\label{sec:ia-sid}

In generative retrieval, SID defines the target space and shapes
query-to-SID learning. Existing SIDs typically rely on static product
semantics, such as titles, descriptions, and categories. Although
these signals preserve product attributes, they do not explicitly
capture search intent or behavioral relationships from user
interactions. To address this limitation, we propose
\emph{Intent-Aware SID Construction} (IA-SID), which integrates
query--product relevance, co-click relationships, and each product's
historical query-intent profile before quantizing the representations
into hierarchical SIDs.

\subsubsection{Query--Product Embedding Alignment}
\label{sec:query-product-alignment}

Constructing search-aligned SIDs requires a representation space that captures behavioral query--product relevance. Yet, general-purpose text encoders are optimized for open-domain semantic similarity and therefore do not explicitly model the implicit relevance signals reflected in e-commerce click logs.

We adapt the general-purpose encoder $f_{\Theta}$ to e-commerce search using historical query--product click logs. Given a clicked pair $(q,i^{+})$ and a set of negative samples $\mathcal{N}_{q}$, we optimize the InfoNCE objective~\cite{oord2019representationlearningcontrastivepredictive}
\begin{equation}
\label{eq:ia-sid-infonce}
\mathcal{L}_{\mathrm{InfoNCE}}
=
-\sum_{(q,i^{+})\in\mathcal{D}}
\log
\frac{
\exp\!\left(\operatorname{sim}(\mathbf{e}_{q},\mathbf{x}_{i^{+}})
  /\tau_{\mathrm{emb}}\right)
}{
\displaystyle
\sum_{i\in\{i^{+}\}\cup\mathcal{N}_{q}}
\exp\!\left(\operatorname{sim}(\mathbf{e}_{q},\mathbf{x}_{i})
  /\tau_{\mathrm{emb}}\right)
}.
\end{equation}
This objective pulls each query toward its clicked product and away from the negative samples. The resulting embeddings are 
$\mathbf{e}_{q}=f_{\Theta}(q)$ and $\mathbf{x}_{i}=f_{\Theta}(t_{i})$, where $t_{i}$ denotes the text associated with product $i$. These embeddings are shared by the subsequent relation-modeling and intent-fusion modules.

\subsubsection{Intent-Guided Product Relation Modeling}
\label{sec:product-relation-modeling}

Query--product contrastive learning captures pairwise relevance between queries and products, but it does not explicitly model inter-product relationships induced by shared queries. To address this limitation, we construct a weighted product-relation graph $G$ from shared-query click behavior, thereby injecting these behavioral relationships into the product representations.

Let $\mathcal{C}_{q}$ denote the set of products clicked under query $q$, and let $n_{i,q}$ denote the cumulative number of clicks received by product $i$ under $q$. We define the edge weight between products $i$ and $j$ as
\begin{equation}
\label{eq:ia-sid-edge-weight}
w_{ij}
=
\sum_{q\in\mathcal{Q}:\,i,j\in\mathcal{C}_{q}}
\sqrt{n_{i,q}n_{j,q}}.
\end{equation}
The geometric mean balances the click evidence for each product pair, while summing over queries accumulates evidence that the products are associated through shared queries. Let $\mathbf{A}_{ij}=w_{ij}$, and define the symmetrically normalized adjacency matrix as $\widehat{\mathbf{A}}=\mathbf{D}^{-\frac{1}{2}}\mathbf{A}
\mathbf{D}^{-\frac{1}{2}}$, where $\mathbf{D}$ is the degree matrix corresponding to $\mathbf{A}$.

Given the search-aligned product representations $\mathbf{X}$, we perform multi-step graph propagation with semantic anchoring:
\begin{equation}
\label{eq:ia-sid-graph-propagation}
\mathbf{H}^{(0)}=\mathbf{X}, \qquad
\mathbf{H}^{(t)}=(1-\alpha_{g})\widehat{\mathbf{A}}\mathbf{H}^{(t-1)}
 +\alpha_{g}\mathbf{X}, \qquad t=1,\ldots,T.
\end{equation}
Neighborhood propagation injects higher-order product relationships induced by shared queries, while the residual term preserves consistency with the original search-aligned representations. Thus, $\mathbf{h}_{i}$ denotes the product representation enriched with co-click structure induced by shared
queries.

\subsubsection{Intent-Enhanced Product Representation Fusion}
\label{sec:intent-fusion}

The query-induced behavioral graph injects search signals through inter-product relationships. Complementarily, we characterize each product using its historical clicked queries to model a query-intent profile, since a single product may match multiple search intents.

For product $i$, we retain the valid query set $\mathcal{Q}_{i}=\left\{q\in\mathcal{Q}\mid n_{i,q}\geq\delta_{c}\right\}$ and aggregate the corresponding query embeddings according to their empirical click frequencies:
\begin{equation}
\label{eq:ia-sid-intent-profile}
a_{i,q}
=
\frac{n_{i,q}}
{\displaystyle\sum_{q'\in\mathcal{Q}_{i}}n_{i,q'}},
\qquad
\mathbf{z}_{i}
=
\sum_{q\in\mathcal{Q}_{i}}a_{i,q}\mathbf{e}_{q}.
\end{equation}
Here, $\mathbf{e}_{q}$ denotes the aligned representation of query $q$. The resulting vector $\mathbf{z}_{i}$ can be viewed as product $i$'s query-intent profile, summarizing the main historical query expressions associated with clicks on the product and their relative frequencies.

Because historical click evidence varies substantially across products, we use confidence-gated residual fusion to control the strength of query-intent injection:
\begin{equation}
\label{eq:ia-sid-fusion}
\widetilde{\mathbf{x}}_{i}
=
\frac{\mathbf{h}_{i}+\lambda\eta_{i}\mathbf{z}_{i}}
{\left\|\mathbf{h}_{i}+\lambda\eta_{i}\mathbf{z}_{i}\right\|_{2}},
\qquad
\eta_{i}
=
\min\!\left(
1,
\frac{\log\!\left(\lvert\mathcal{Q}_{i}\rvert+1\right)}
     {\log\!\left(Q_{\max}+1\right)}
\right).
\end{equation}
Here, $\lambda$ controls the fusion strength, $\eta_{i}$ is a product-specific confidence gate derived from the number of retained queries, and $Q_{\max}$ is the saturation threshold for this gate. The gate prevents sparse or noisy query profiles from dominating product representations, while allowing products with sufficient historical query--click evidence to receive stronger intent signals.

We quantize the enhanced representation $\widetilde{\mathbf{x}}_{i}$
using an $L$-stage residual $K$-means quantizer (RQ--KMeans)~
\cite{10.1145/3705328.3759300}, and concatenate the codes from stages
to form product $i$'s hierarchical SID
$\mathbf{s}_{i}=\left(c_{i}^{(1)},c_{i}^{(2)},\ldots,c_{i}^{(L)}\right)$.
This process maps each product to an SID sequence that encodes
search-intent structure. The SID serves as the prediction target for
the generative model, enabling query-to-SID learning in a target space
more closely aligned with search behavior.


\begin{table*}[t]
  \caption{Examples of multi-granularity query synthesis in SQE-SFT.}
  \label{tab:multi-granularity-query-synthesis}
  \centering
  \small

  \begin{tabular}{@{}p{0.27\textwidth}p{0.67\textwidth}@{}}
    \toprule
    \textbf{Stage / Query Type} & \textbf{Example Output} \\
    \midrule

    \textbf{Raw title} ($t_i$) &
    JBL TUNE BUDS2 second-generation Liulidou true wireless
    Bluetooth earbuds, featuring an in-ear design, call noise
    reduction, dual-ear transmission, compatibility with Huawei
    and Apple smartphones, and an upgraded white finish \\
    \addlinespace[4pt]

    \textbf{Normalized semantics} ($\mathcal{E}(t_i)$) &
    JBL TUNE BUDS2 second-generation Liulidou true wireless
    earbuds in white, featuring an in-ear design, call noise
    reduction, dual-ear transmission, and compatibility with
    Huawei and Apple smartphones \\

    \midrule

    \textbf{Product identity} ($\mathcal{Q}_i^{\mathrm{id}}$) &
    \textit{(i)} JBL TUNE BUDS2;\newline
    \textit{(ii)} JBL TUNE BUDS2 true wireless earbuds;\newline
    \textit{(iii)} JBL second-generation Liulidou earbuds \\
    \addlinespace[4pt]

    \textbf{Category scope} ($\mathcal{Q}_i^{\mathrm{cat}}$) &
    JBL true wireless earbuds \\
    \addlinespace[4pt]

    \textbf{Series/variant} ($\mathcal{Q}_i^{\mathrm{var}}$) &
    JBL TUNE BUDS \\
    \addlinespace[4pt]

    \textbf{Attribute constraints} ($\mathcal{Q}_i^{\mathrm{attr}}$) &
    \textit{(i)} JBL TUNE BUDS2 in-ear earbuds;\newline
    \textit{(ii)} JBL TUNE BUDS2 earbuds with call noise reduction
    and dual-ear transmission;\newline
    \textit{(iii)} white JBL TUNE BUDS2 earbuds compatible with
    Huawei and Apple smartphones \\

    \bottomrule
  \end{tabular}
\end{table*}

\subsection{Synthetic Query-Enhanced Unified SFT}
\label{sec:sqe-sft}

Even when SIDs encode query intent, the generative model must learn to map natural-language queries to the corresponding SIDs. SFT formulations combine product--SID association learning with query alignment. Product-to-SID and SID-to-product tasks teach the model to associate catalog items with SIDs, but their inputs differ from the natural-language queries encountered at inference time. Direct query-to-SID training on interaction logs is better aligned with the target task, yet provides limited coverage for long-tail products because such queries are sparse. We therefore propose \emph{Synthetic Query-Enhanced Unified SFT} (SQE-SFT), which generates multi-granularity synthetic queries from catalog information and jointly trains them with real interaction queries under a unified query-to-SID objective. This design expands supervision coverage while preserving alignment with the observed query distribution.

\subsubsection{Multi-Granularity Synthetic Query Construction}
\label{sec:multi-granularity-synthetic-query}

Given the catalog text $t_{i}$ of product $i$, we first use a product semantic parsing model to extract its structured semantics:
\begin{equation}
\label{eq:sqe-semantic-parsing}
\mathcal{E}(t_{i})
=
\left(
\mathrm{spu}_{i},
\mathcal{B}_{i},
\mathcal{P}_{i},
\mathcal{R}_{i},
\mathcal{M}_{i},
\mathcal{A}_{i}
\right).
\end{equation}
Here, $\mathrm{spu}_{i}$ denotes the normalized product name; $\mathcal{B}_{i}$, $\mathcal{P}_{i}$, $\mathcal{R}_{i}$, and $\mathcal{M}_{i}$ denote the sets of brands, categories, series, and models, respectively; and $\mathcal{A}_{i}$ denotes the set of attributes. We use these fields to model two common retrieval behaviors: specifying a target product through identity or hierarchical semantics, and constraining a product query with explicit
attributes.

\paragraph{Entity-Specifying Queries.}
Users may specify a target product directly by name or narrow the candidate product scope using its category, brand, series, or model. To cover these behaviors, we construct queries from three complementary views: product identity, category scope, and product variant, denoted by $\mathcal{G}=\{\mathrm{id},\mathrm{cat},\mathrm{var}\}$. For each view $g\in\mathcal{G}$, we combine the corresponding fields of product $i$ into a candidate semantic set $\mathcal{V}_{i}^{g}$. The resulting entity-specifying query set is
\begin{equation}
\label{eq:sqe-entity-query-set}
\mathcal{Q}_{i}^{\mathrm{ent}}
=
\left\{
\tau_{\mathrm{ent}}(g,v)
\mid
g\in\mathcal{G},\;
v\in\mathcal{V}_{i}^{g}
\right\},
\end{equation}
where $\tau_{\mathrm{ent}}(g,\cdot)$ verbalizes the structured semantic combination under view $g$ as a natural-language query.

\paragraph{Attribute-Constraining Queries.}
To cover retrieval intents driven by product attributes or usage requirements, we first discard attributes in $\mathcal{A}_{i}$ that occur too infrequently across the catalog, yielding the filtered attribute set $\mathcal{A}_{i}^{+}$. We then use the normalized product name $\mathrm{spu}_{i}$ as the query anchor to retain its association with the target product, and sample attribute subsets to construct attribute-constraining queries:
\begin{equation}
\label{eq:sqe-attribute-query-set}
\mathcal{Q}_{i}^{\mathrm{attr}}
=
\left\{
\tau_{\mathrm{attr}}
\left(\mathrm{spu}_{i},\widetilde{\mathcal{A}}_{i}\right)
\;\middle|
\widetilde{\mathcal{A}}_{i}
\in
\operatorname{Sample}\!\left(\mathcal{A}_{i}^{+}\right)
\right\}.
\end{equation}
Here, $\operatorname{Sample}(\cdot)$ generates nonempty attribute subsets of varying sizes, and $\tau_{\mathrm{attr}}(\cdot,\cdot)$ verbalizes the product name and selected attributes as a natural-language query.

Finally, we merge the two query sets and apply normalization, deduplication, and validity filtering to obtain the synthetic query set $\mathcal{Q}_{i}^{\mathrm{syn}}$ for product $i$. Table~\ref{tab:multi-granularity-query-synthesis} presents examples generated from a real catalog item.

\subsubsection{Unified Two-Stage SFT}
\label{sec:unified-two-stage-sft}

Let $\mathcal{Q}_{i}^{\mathrm{real}}$ denote the set of real user queries associated with clicks or conversions on product $i$. For any $q\in\mathcal{Q}_{i}^{\mathrm{syn}}\cup\mathcal{Q}_{i}^{\mathrm{real}}$, we use the following unified instruction template:
\begin{center}
\parbox{0.96\columnwidth}{
\centering
\textnormal{``Given the user request `}$q$
\textnormal{', recommend the most relevant SID.''}
}
\end{center}
The desired response is the SID $\mathbf{s}_{i}$ of the product associated with $q$. In the first stage, we train the model on synthetic query-to-SID examples to learn the catalog-wide mapping from product semantics to SIDs. This stage provides direct supervision for long-tail products and attribute-based intents. We then fine-tune the model on real query-to-SID pairs to align it with the distribution of real user queries and observed interactions. Both stages optimize the same query-to-SID objective, and the resulting model serves as the initialization for subsequent preference optimization.

\subsection{Relevance-Calibrated Preference Optimization}
\label{sec:rcpo}

Although SFT teaches the model to map queries to relevant SIDs, it does
not explicitly model business preferences reflected in user behavior and
commercial outcomes. Directly optimizing these business signals without
relevance constraints may favor commercially valuable products that do
not match the query intent. We therefore introduce
\emph{Relevance-Calibrated Preference Optimization} (RCPO), which
integrates business preference learning with relevance-aware preference
construction. RCPO restricts preference comparisons to semantically
relevant candidates, uses behavioral and business-value signals to refine
their ordering, and adapts the weight of each preference update to the
resulting pairwise composite-score margin. This design enables
business-aware preference learning while preserving query-intent
consistency.

\subsubsection{Multi-Signal Preference Pair Construction}
\label{sec:multi-signal-preference-pairs}

Given a query $q$, beam search with the generative retrieval model yields
candidate SIDs
$\mathcal{R}_{q}=\{\hat{\mathbf{s}}_{k}\}_{k=1}^{N_{\mathrm{beam}}}$.
Let $\mathcal{H}_{q}^{\mathrm{sid}}$ denote SIDs receiving positive
feedback for $q$ in the interaction logs. We measure semantic relevance
by mapping each candidate SID $\hat{\mathbf{s}}$ to its natural-language
description $\mathbf{t}(\hat{\mathbf{s}})$ and scoring the resulting
query--description pair with a fine-tuned scorer $g_{\phi}$:
$
\mathrm{SRS}(q,\hat{\mathbf{s}})
=g_{\phi}\bigl(q,\mathbf{t}(\hat{\mathbf{s}})\bigr)
$.

Because semantic relevance alone does not capture user preferences among
multiple relevant items, we further estimate the \emph{Smoothed Business
Preference} (SBP) from the interaction logs. SBP aggregates click, payment,
and transaction-value signals and applies category-wise mean smoothing to
mitigate estimation bias from sparse interactions. Its full definition is
provided in Appendix~\ref{app:sbp}.

Based on SRS and SBP, we construct preference pairs using two complementary strategies. \emph{Semantic Relevance Anchoring} (SRA) determines the preference direction according to semantic relevance, whereas \emph{Business Preference Refinement} (BPR) refines the business-preference ordering among semantically comparable candidates. For each query with historical positive interactions, SRA selects the highest-SRS SID in $\mathcal{H}_{q}^{\mathrm{sid}}$ as the preferred SID and a relatively low-SRS candidate from $\mathcal{R}_{q} \setminus \mathcal{H}_{q}^{\mathrm{sid}}$ as the rejected SID. BPR compares candidates for the same query that satisfy the relevance constraint and have an SRS gap no larger than a predefined margin, preferring the candidate with the higher SBP over the other. The resulting preference pairs constitute the offline dataset $\mathcal{D}_{\mathrm{pref}} = \{(q, \hat{\mathbf{s}}^{+}, \hat{\mathbf{s}}^{-})\}$.

\subsubsection{RCPO Objective}
\label{sec:relevance-calibrated-dpo}

Given $\mathcal{D}_{\mathrm{pref}}$, we initialize the policy model
$\pi_{\theta}$ from the SQE-SFT model and freeze a copy as the reference model
$\pi_{\mathrm{ref}}$. Although relevance-constrained preference construction
yields semantically comparable pairs, it does not quantify how strongly
offline evidence supports the assigned preference within each pair. We
therefore compute an offline composite score for each query--SID pair as the
equal-weight average of its SRS and SBP, both normalized to $[0,1]$.

For each preference pair, let $r^{+}$ and $r^{-}$ denote the composite scores
of the preferred and rejected SIDs, respectively, and define the signed
calibration margin as $\Delta_{r}=r^{+}-r^{-}$. We use this margin to determine
the pair-specific RCPO coefficient:
\begin{equation}
\label{eq:rcpo-effective-beta}
\beta_{\mathrm{eff}}
=\beta\cdot
\operatorname{clip}\!\left(
1+\log\frac{2-\Delta_{r}}{2+\Delta_{r}},\,0.2,\,1.8
\right).
\end{equation}
Here, $\beta$ is the base temperature. The coefficient
decreases as $\Delta_{r}$ increases: a large positive margin indicates strong
offline support for the preferred SID, so we place less emphasis on such
pairs to avoid redundant optimization. Smaller margins receive larger
coefficients, increasing the influence of less decisive preference pairs in
the RCPO objective. If either score is unavailable, we set the multiplicative
calibration factor to $1$.

For a preference pair
$(\hat{\mathbf{s}}^{+},\hat{\mathbf{s}}^{-})$, we define the relative
preference difference between the policy and reference models as
\begin{equation}
\label{eq:rcpo-relative-preference}
\Delta_{\theta}
=
\log\frac{
\pi_{\theta}(\hat{\mathbf{s}}^{+}\mid x_{q})
}{
\pi_{\mathrm{ref}}(\hat{\mathbf{s}}^{+}\mid x_{q})
}
-
\log\frac{
\pi_{\theta}(\hat{\mathbf{s}}^{-}\mid x_{q})
}{
\pi_{\mathrm{ref}}(\hat{\mathbf{s}}^{-}\mid x_{q})
}.
\end{equation}
We then optimize the following RCPO objective with a length-normalized
auxiliary SFT term for the preferred SID:
\begin{equation}
\label{eq:rcpo-objective}
\mathcal{L}
=-\mathbb{E}_{\mathcal{D}_{\mathrm{pref}}}
\Biggl[
\log\sigma\!\left(\beta_{\mathrm{eff}}\Delta_{\theta}\right)
+
\frac{\eta_{\mathrm{sft}}}{\lvert\hat{\mathbf{s}}^{+}\rvert}
\log\pi_{\theta}(\hat{\mathbf{s}}^{+}\mid x_{q})
\Biggr].
\end{equation}
Here, $\eta_{\mathrm{sft}}$ controls the weight of the auxiliary SFT term and
$x_{q}$ denotes the query input. Together, relevance-constrained preference
construction and RCPO enable the model to learn business preference while preserving alignment with query intent.

\section{Experiments}


\begin{table*}[t]
  \caption{Main offline retrieval results under cumulative component additions.}
  \label{tab:offline-retrieval-results}
  \centering
  \small
  \setlength{\tabcolsep}{6pt}
  \begin{tabular}{@{}lcccccc@{}}
    \toprule
    \textbf{Method}
      & \textbf{Recall@20}
      & \textbf{MRR@20}
      & \textbf{NDCG@20}
      & \textbf{Recall@50}
      & \textbf{MRR@50}
      & \textbf{NDCG@50} \\
    \midrule
    ProdGR          & 0.4247 & 0.1982 & 0.2318 & 0.5606 & 0.2022 & 0.2623 \\
    +SQE-SFT        & 0.4954 & 0.2381 & 0.2784 & 0.6255 & 0.2418 & 0.3079 \\
    +IA-SID         & 0.5006 & 0.2433 & 0.2841 & 0.6312 & 0.2471 & 0.3136 \\
    \textbf{+RCPO (ICEGR)}
      & \textbf{0.5169}
      & \textbf{0.2509}
      & \textbf{0.2935}
      & \textbf{0.6506}
      & \textbf{0.2548}
      & \textbf{0.3237} \\
    \bottomrule
  \end{tabular}
\end{table*}

\subsection{Experimental Setup}
\label{sec:experimental-setup}

\paragraph{Dataset.}
We chronologically split nearly three months of Baidu e-commerce search logs,
reserving the final day for testing and using the remaining data for training
and validation. The logs cover $12.8\,\mathrm{M}$ distinct queries and
$14.6\,\mathrm{M}$ distinct products. All training artifacts, including SID
construction and sample generation, are derived exclusively from pre-test data
to prevent temporal leakage.

\paragraph{Offline Comparison and Ablation Settings.}
We use ProdGR, the production generative retrieval model, as the
baseline. Because prior industrial generative retrieval systems are
trained and served with proprietary data and infrastructure, their
results are not directly comparable with ours. We therefore compare
ProdGR and ICEGR under the same 0.5B backbone, RQ--KMeans SID
construction, data split, SFT/RL training setting, and beam-search
configuration; each ablation changes only the component under study.

\paragraph{Evaluation Metrics.}
For each test query, we take the set of products the user interacted with
as ground truth and evaluate the retrieved list with Recall@$K$, MRR@$K$
and NDCG@$K$. Relevance is binary, MRR uses the rank of the first relevant
product, NDCG is normalized by the ideal ranking truncated at $K$, and all
metrics are macro-averaged over queries.

\paragraph{Implementation Details.}
We use an 8B e-commerce-adapted embedding model, producing
256-dimensional fused representations. In IA-SID, we perform $T=3$
graph propagation steps with $\alpha_g=0.2$. Product SIDs are generated using a three-level RQ--KMeans quantizer with 1,024 codes
each, where $Q_{\max}=50$ and $\lambda=1.0$. Preference pairs are
constructed using a 0.6B semantic relevance model.

\subsection{Offline Performance}
\label{sec:offline-performance}
For offline evaluation, we sample 20{,}000 distinct queries from the held-out
final day of user search logs and use their clicked products as ground-truth
relevant items. Both ICEGR and ProdGR are evaluated on the same query set.
ICEGR consistently outperforms ProdGR at both cutoffs (Table~\ref{tab:offline-retrieval-results}).
At $K{=}20$, Recall@$20$, MRR@$20$, and NDCG@$20$ improve by 21.7\%, 26.6\%,
and 26.6\%, respectively; the corresponding gains at $K{=}50$ are 16.1\%,
26.0\%, and 23.4\%. The larger gains in MRR and NDCG indicate that ICEGR
improves the ordering of relevant items near the top of the retrieved list,
beyond increasing relevant-item coverage.



\subsection{Ablation Study}
\label{sec:ablation-study}

\paragraph{Overall component contributions.}
We first quantify the contributions of ICEGR's three query-intent-preserving
stages. Under the component-addition protocol in
Table~\ref{tab:offline-retrieval-results}, SQE-SFT accounts for 76.7\% of the
cumulative Recall@$20$ gain, followed by RCPO (17.7\%) and IA-SID (5.6\%).
This decomposition suggests that expanded query-to-SID supervision is the
dominant source of the observed improvement. SQE-SFT is further analyzed
across product popularity levels in Section~\ref{sec:synthetic-supervision}.

\paragraph{IA-SID}
Replacing the search-aligned encoder with the base encoder reduces
Recall@$20$ from 0.5169 to 0.4179, below ProdGR's 0.4247
(Table~\ref{tab:ia-sid-ablation}), showing that generic semantic
representations alone are insufficient to capture retrieval-specific
query--product relevance. The two IA-SID mechanisms have distinct effects:
removing intent-guided product-relation modeling lowers Recall@$20$ by 4.35\%,
whereas removing intent-enhanced product-representation fusion lowers
MRR@$20$ by 7.21\%. The former contributes more to relevant-item coverage,
while the latter more strongly affects the rank of the first relevant item.
Removing both mechanisms causes a 13.60\% drop in Recall@$20$, exceeding the
9.13\% sum of their individual drops and indicating complementary search
signals.

\begin{figure}[htbp]
  \centering
  \includegraphics[
    width=\columnwidth
  ]{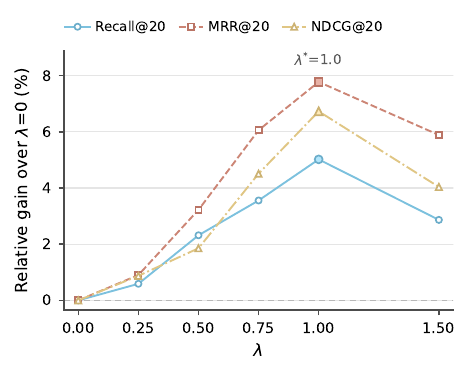}
  \caption{Sensitivity analysis of the query-intent fusion weight
  $\lambda$.}
  \Description{Sensitivity of model performance to the query-intent fusion
  weight lambda. The results show how retrieval performance changes as the
  fusion weight varies.}
  \label{fig:fusion-weight-sensitivity}
\end{figure}

\paragraph{Fusion-weight sensitivity}
Performance on the validation set follows an inverted-U-shaped trend as the
global fusion weight $\lambda$ increases, peaking at $\lambda=1.0$
(Figure~\ref{fig:fusion-weight-sensitivity}). We use this value in all
subsequent experiments. The trend indicates that effective fusion must balance
query-intent signals with product semantics.


\begin{table}[htbp]
  \caption{Ablation study of IA-SID at cutoff 20.}
  \label{tab:ia-sid-ablation}
  \centering
  \footnotesize
  \setlength{\tabcolsep}{2pt}

  \begin{tabular}{@{}p{0.40\columnwidth}ccc@{}}
    \toprule
    \textbf{Method}
      & \textbf{Recall@20}
      & \textbf{MRR@20}
      & \textbf{NDCG@20} \\
    \midrule
    \raggedright\textbf{ICEGR}
      & \textbf{0.5169}
      & \textbf{0.2509}
      & \textbf{0.2935} \\
    \raggedright w/o Query--Product Alignment
      & 0.4179~(-19.15\%)
      & 0.2154~(-14.15\%)
      & 0.2449~(-16.56\%) \\
    \raggedright w/o Relation Modeling
      & 0.4944~(-4.35\%)
      & 0.2490~(-0.75\%)
      & 0.2857~(-2.67\%) \\
    \raggedright w/o Representation Fusion
      & 0.4922~(-4.78\%)
      & 0.2328~(-7.21\%)
      & 0.2750~(-6.30\%) \\
    \raggedright w/o Both Modules
      & 0.4466~(-13.60\%)
      & 0.2314~(-7.78\%)
      & 0.2605~(-11.24\%) \\
    \bottomrule
  \end{tabular}
\end{table}

\paragraph{RCPO}
We assess whether RCPO can learn business preferences while preserving query
relevance. Training on preferred SIDs lowers Recall@$20$ by 2.46\%
relative to ICEGR, showing that the improvement does not arise merely from
adding high-quality query-to-SID pairs
(Table~\ref{tab:rcpo-ablation-control}). Reversing the preference labels
reduces MRR@$20$ to 0.1719, below ProdGR's 0.1982, confirming the
importance of the preference direction. With preference pairs and settings fixed, replacing pair-specific $\beta_{\mathrm{eff}}$ with a fixed
$\beta$ lowers NDCG@$20$ by 7.36\%, whereas removing SBP most strongly
reduces MRR@$20$ (6.26\%). The two mechanisms play complementary roles:
pair-specific calibration improves ranking quality across the top results,
while SBP more strongly affects the rank of the first relevant item.

\begin{figure*}[htbp]
  \centering
  \includegraphics[
    width=\textwidth
  ]{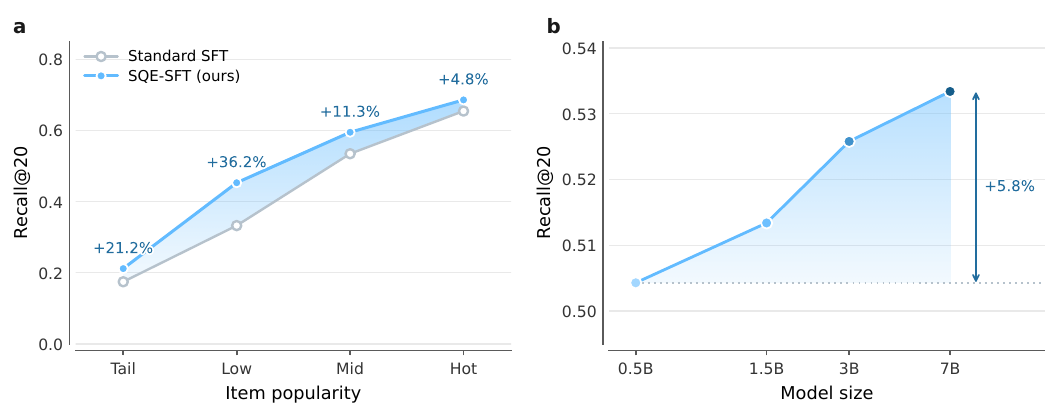}
  \caption{Further analysis of ICEGR: (a) SQE-SFT gains across
item-popularity groups; (b) Recall@$20$ scaling with backbone size.}
  \Description{Two-panel analysis of ICEGR retrieval performance.
Panel (a) compares SQE-SFT gains across item-popularity groups.
Panel (b) shows Recall@$20$ for different backbone sizes.}
  \label{fig:retrieval-bottlenecks}
\end{figure*}

\paragraph{Human relevance evaluation}
We conduct a blinded human evaluation of anonymized top-five results
for 50 randomly sampled test queries. The relevance rate decreases
from 0.87 with ICEGR to 0.58 with fixed $\beta$. Representative cases in
Appendix~\ref{app:qualitative-examples} show that the fixed-$\beta$ variant
can promote business-preferred yet query-irrelevant products to the top ranks,
demonstrating the role of pair-specific DPO calibration in preserving
query--product relevance.


\begin{table}[htbp]
  \caption{Ablation and control study of RCPO at cutoff 20.}
  \label{tab:rcpo-ablation-control}
  \centering
  \footnotesize
  \setlength{\tabcolsep}{1pt}

  \newcommand{\tablefourmetric}[2]{%
    \begin{tabular}[c]{@{}c@{}}#1\\[-1pt](#2)\end{tabular}%
  }

  \begin{tabular}{@{}p{0.38\columnwidth}@{\hspace{4pt}}cccc@{}}
    \toprule
    \textbf{Method}
      & \textbf{Recall@20}
      & \textbf{MRR@20}
      & \textbf{NDCG@20}
      & \textbf{Relevance} \\
    \midrule
    \parbox[c]{0.38\columnwidth}{\raggedright\textbf{ICEGR}}
      & \textbf{0.5169}
      & \textbf{0.2509}
      & \textbf{0.2935}
      & \textbf{0.87} \\
    \parbox[c]{0.38\columnwidth}{\raggedright w/o Dynamic $\beta_{\mathrm{eff}}$}
      & \tablefourmetric{0.4937}{-4.49\%}
      & \tablefourmetric{0.2372}{-5.46\%}
      & \tablefourmetric{0.2719}{-7.36\%}
      & \tablefourmetric{\textbf{0.58}}{\textbf{-33.3\%}} \\
    \parbox[c]{0.38\columnwidth}{\raggedright Preferred-only SFT}
      & \tablefourmetric{0.5042}{-2.46\%}
      & \tablefourmetric{0.2420}{-3.55\%}
      & \tablefourmetric{0.2858}{-2.62\%}
      & -- \\
    \parbox[c]{0.38\columnwidth}{\raggedright Preference Label Reversed}
      & \tablefourmetric{0.4192}{-18.90\%}
      & \tablefourmetric{0.1719}{-31.49\%}
      & \tablefourmetric{0.2152}{-26.68\%}
      & -- \\

    \parbox[c]{0.38\columnwidth}{\raggedright w/o Business Preference (SBP)}
      & \tablefourmetric{0.5016}{-2.96\%}
      & \tablefourmetric{0.2352}{-6.26\%}
      & \tablefourmetric{0.2798}{-4.67\%}
      & -- \\
    \bottomrule
  \end{tabular}
\end{table}


\subsection{Further Analysis}
\label{sec:further-analysis}
We study how query-to-SID supervision and backbone capacity affect
ICEGR retrieval quality.

\subsubsection{Synthetic Supervision Benefits Less Popular Items}
\label{sec:synthetic-supervision}
To isolate the contribution of SQE-SFT, we compare ICEGR with a variant that replaces SQE-SFT with ProdGR's SFT procedure while keeping IA-SID and RCPO unchanged. We partition test query--item pairs into Tail, Low, Mid, and Hot groups according to the target item's training-set click frequency, with thresholds given in Appendix~\ref{app:popularity-evaluation}.

As shown in Figure~\ref{fig:retrieval-bottlenecks}(a), SQE-SFT improves performance in four groups, with the largest Recall@$20$ gain in \textbf{Low (36.2\%)}, followed by Tail (21.2\%), and progressively smaller gains in Mid and Hot. MRR@$20$ and NDCG@$20$ show similar trends (Appendix Table~\ref{tab:popularity-wise-performance}). The decline from Low to Hot is consistent with diminishing returns from synthetic supervision as real supervision becomes more abundant: Mid and Hot items already receive substantial interaction-based supervision, leaving less room for synthetic queries.

Interestingly, Low achieves a larger gain than Tail despite having denser real supervision. This result suggests that synthetic and real supervision are complementary. Tail items rely more heavily on synthetic-query coverage, but their gains may be limited by a mismatch between synthetic and real-user query distributions. Low items benefit from both broad synthetic coverage and enough real interaction data to preserve alignment with real-user queries, yielding the largest improvement. These findings support the unified two-stage SFT design.

\subsubsection{Scaling Law Verification}
\label{sec:backbone-scaling}

We fix the training data, SID codebooks, and decoding configuration,
and vary only the backbone size among 0.5B, 1.5B, 3B, and 7B. For each,
we report the checkpoint with the best validation performance
over five SFT epochs. Figure~\ref{fig:retrieval-bottlenecks}(b) shows
that Recall@$20$ increases monotonically from
$0.5043$ at 0.5B to $0.5334$ at 7B. From 0.5B to 7B, MRR@$20$ and
NDCG@$20$ increase by $5.1\%$ and $5.8\%$, respectively. These results
demonstrate the availability of LLM Scaling Law for ICEGR, with no clear
evidence of a performance plateau within the evaluated range.

Together, these results show that supervision coverage and backbone
capacity are complementary sources of improvement. SQE-SFT is
particularly beneficial for less popular items, while backbone scaling
continues to improve retrieval quality across the evaluated range. We
therefore deploy the 0.5B model online, which provides a strong
accuracy--efficiency trade-off.


\begin{table*}[t]
  \caption{Online A/B testing results of ICEGR relative to the production MCA baseline.}
  \label{tab:online-ab-testing}
  \centering
  \small
  \setlength{\tabcolsep}{2pt}

  \newcommand{\tablefiveheader}[1]{%
    \parbox[c][4.5ex][c]{\linewidth}{\centering #1}%
  }
  \newcommand{\tablefivevalue}[1]{%
    \parbox[c][2.8ex][c]{\linewidth}{\centering\bfseries #1}%
  }

  \begin{tabular}{@{}
      p{0.10\textwidth}
      p{0.08\textwidth}
      p{0.14\textwidth}
      p{0.20\textwidth}
      p{0.14\textwidth}
      p{0.10\textwidth}@{}}
    \toprule
    \multicolumn{1}{c}{\textit{Exposure}}
      & \multicolumn{3}{c}{\textit{User Engagement}}
      & \multicolumn{2}{c}{\textit{Business Conversion}} \\
    \cmidrule(lr){1-1}
    \cmidrule(lr){2-4}
    \cmidrule(lr){5-6}
    \tablefiveheader{Show PV}
      & \tablefiveheader{CTR}
      & \tablefiveheader{Long-click Rate}
      & \tablefiveheader{Product Landing Page PV}
      & \tablefiveheader{Order Volume}
      & \tablefiveheader{GMV} \\
    \midrule
    \tablefivevalue{+8.84\%}
      & \tablefivevalue{+3.52\%}
      & \tablefivevalue{+5.27\%}
      & \tablefivevalue{+8.75\%}
      & \tablefivevalue{+15.96\%}
      & \tablefivevalue{+7.53\%} \\
    \bottomrule
  \end{tabular}

  \begin{minipage}{\textwidth}
    \footnotesize
    \emph{Note.} Show PV denotes the exposed number of products in Baidu Search result page. CTR is the click-through rate, calculated as the number of product clicks divided by Show PV. Long-click
    Rate measures post-click engagement based on long-click behavior. Product
    Landing Page PV denotes the number of page views generated when users jump from Baidu search result to a product landing page(may belongs to Jingdong, Taobao or other E-commerce platforms)
  \end{minipage}
\end{table*}

\subsection{Online A/B Testing}
\label{sec:online-ab-testing}
We deployed the 0.5B-parameter ICEGR model in Baidu E-commerce Search system and ran a two-week A/B test on 20\% of live traffic. The control group used the existing MCA system, whereas the treatment group used the same production pipeline augmented with ICEGR. A fixed number of ICEGR's top-$k$ results was inserted directly into the final impression list, creating an end-to-end generative retrieval path from user queries to product impressions. The remaining candidates entered the existing relevance-filtering and ranking pipeline. For confidentiality reasons, we report only relative improvements over the control group.

As shown in Table~\ref{tab:online-ab-testing}, ICEGR increases Show PV, CTR, and Long-click Rate by \textbf{8.84\%}, \textbf{3.52\%}, and \textbf{5.27\%}, respectively. Product Landing Page PV also increases by \textbf{8.75\%}, while Order Volume and GMV rise by \textbf{15.96\%} and \textbf{7.53\%}, respectively. These results show that ICEGR expands product exposure while improving click efficiency, post-click engagement, and business outcomes.

\subsection{Online Serving Efficiency}
\label{sec:online-serving-efficiency}
The computational cost of autoregressive decoding is a major challenge in deploying generative retrieval systems at scale. ICEGR runs on a cluster of 67 NVIDIA L20 GPUs and, with a beam size of 50, sustains a peak load of $2{,}200$ QPS with an average latency of $156\,\mathrm{ms}$, meeting the throughput and latency requirements of the current production environment.

\section{Conclusion}

We introduced ICEGR, an intent-coherent generative retrieval framework
for e-commerce search. Its central design principle is to preserve
query intent throughout the retrieval pipeline, from semantic-ID
construction and query-to-SID supervision to preference optimization.
ICEGR improves $\mathrm{Recall}@20$ by 21.7\% and
$\mathrm{NDCG}@20$ by 26.6\% over the production generative retrieval
baseline. Our analysis further shows that improving query-to-SID
supervision is especially important for less popular items, while
increasing backbone capacity provides consistent additional gains
within the evaluated range. RCPO complements these improvements by
introducing business-aware preferences only among semantically
relevant candidates and adapting the optimization strength to the
preference margin. Online A/B testing confirms that ICEGR improves CTR,
order volume, and GMV while satisfying production latency and
throughput requirements.

\appendix

\section{Construction of the Smoothed Business Preference (SBP)}
\label{app:sbp}

For impression $i$ of $(q,s)$, clicks, payments, and transaction value
capture attractiveness, conversion, and monetary value, respectively. To
mitigate high-value outliers, we log-scale and clip $g_i$ as
\begin{equation}
\label{eq:sbp-clipped-gmv}
\bar{g}_i =
\min\!\left(
  \frac{\log(1+g_i)}{\log(1+g_{0.95})},
  1
\right),
\end{equation}
where $g_i$ is the transaction value, set to zero if no payment occurs, and
$g_{0.95}$ is the 95th percentile of nonzero transaction values in the
training set. We define the impression-level business reward as
\begin{equation}
\label{eq:sbp-impression-reward}
r_i =
\lambda_{\mathrm{clk}} y_i^{\mathrm{clk}}
 + \lambda_{\mathrm{pay}} y_i^{\mathrm{pay}}
 + \lambda_{\mathrm{gmv}} \bar{g}_i,
\end{equation}
where $y_i^{\mathrm{clk}}$ and $y_i^{\mathrm{pay}}$ are binary click and
payment indicators, respectively, and the three weights sum to one.

For each query--product pair $(q,s)$, we aggregate its historical rewards
using category-level smoothing:
\begin{equation}
\label{eq:sbp-definition}
\mathrm{SBP}(q,s) =
\frac{
  \displaystyle\sum_{i\in\mathcal{I}_{q,s}} r_i
  + \kappa\mu_{c(s)}
}{
  \lvert\mathcal{I}_{q,s}\rvert + \kappa
},
\end{equation}
where $\mathcal{I}_{q,s}$ is the set of historical impressions of $(q,s)$,
$c(s)$ denotes $s$'s category, and $\mu_{c(s)}$ its mean training-set reward.
The parameter $\kappa$ controls prior strength and reduces variance for sparse
pairs.
We set
$\lambda_{\mathrm{clk}}=0.2$,
$\lambda_{\mathrm{pay}}=0.3$,
$\lambda_{\mathrm{gmv}}=0.5$, and $\kappa=10$.
These values were selected based on downstream validation performance.

\section{Qualitative Analysis of DPO Calibration}
\label{app:qualitative-examples}
ICEGR achieved a higher top-five relevance rate on 32 queries and tied on 18.
Table~\ref{tab:representative-top5-results} presents two examples. For
``Omega repair center,'' the fixed-$\beta$ variant ranks Omega watches, whereas
ICEGR returns five repair services. For ``Which Zongshen enclosed electric
tricycle is good,'' the fixed-$\beta$ variant mixes gasoline and electric
tricycles, whereas ICEGR returns only enclosed electric tricycles.


\begin{table}[t]
  \caption{Top-5 results for two representative queries. \protect\ding{51} and
  \protect\ding{55} indicate human-annotated relevance. Titles are translated
  and abbreviated.}
  \label{tab:representative-top5-results}
  \centering

  {\footnotesize

  \newcommand{\tablesixcell}[1]{%
    \parbox[t]{0.40\columnwidth}{\raggedright #1}%
  }

  \begin{tabular}{@{}c@{\hspace{0.25em}}c@{\hspace{0.15em}}c@{}}
    \toprule
    \textbf{Rank} & \textbf{Fixed-$\beta$ DPO} & \textbf{ICEGR} \\
    \midrule

    \multicolumn{3}{@{}l}{\textit{(a) Query: ``Omega repair center''}} \\
    \addlinespace[2pt]
    1 & \tablesixcell{Seamaster 300 watch \ding{55}}
      & \tablesixcell{Repair: cleaning \& battery \ding{51}} \\
    \addlinespace[1pt]
    2 & \tablesixcell{De Ville watch \ding{55}}
      & \tablesixcell{Repair: movement oiling \ding{51}} \\
    \addlinespace[1pt]
    3 & \tablesixcell{Repair: cleaning \& battery \ding{51}}
      & \tablesixcell{Repair: battery replacement \ding{51}} \\
    \addlinespace[1pt]
    4 & \tablesixcell{De Ville watch, variant \ding{55}}
      & \tablesixcell{Repair: appointment service \ding{51}} \\
    \addlinespace[1pt]
    5 & \tablesixcell{Constellation 18K watch \ding{55}}
      & \tablesixcell{Repair: refinishing \& polishing \ding{51}} \\

    \midrule
    \multicolumn{3}{@{}l}{\textit{(b) Query: ``Which Zongshen enclosed electric tricycle is good''}} \\
    \addlinespace[2pt]
    1 & \tablesixcell{Enclosed electric, 5-door \ding{51}}
      & \tablesixcell{Enclosed electric, 5-door \ding{51}} \\
    \addlinespace[1pt]
    2 & \tablesixcell{Gasoline cargo tipper \ding{55}}
      & \tablesixcell{Enclosed electric, 4-door \ding{51}} \\
    \addlinespace[1pt]
    3 & \tablesixcell{Gasoline flat-seat tricycle \ding{55}}
      & \tablesixcell{Enclosed electric, 60V full cab \ding{51}} \\
    \addlinespace[1pt]
    4 & \tablesixcell{Enclosed electric, 4-door \ding{51}}
      & \tablesixcell{Enclosed electric, family-use \ding{51}} \\
    \addlinespace[1pt]
    5 & \tablesixcell{Gasoline 200, water-cooled \ding{55}}
      & \tablesixcell{Enclosed electric, 60V/1500W \ding{51}} \\

    \bottomrule
  \end{tabular}
  }
\end{table}

\section{Popularity-Stratified Evaluation of SQE-SFT}
\label{app:popularity-evaluation}

For each test query--item pair, let $c_i$ be the target item's
training-set click count. We partition the pairs into four mutually
exclusive buckets: Tail ($c_i \leq 2$), Low ($2 < c_i \leq 9$),
Mid ($9 < c_i \leq 99$), and Hot ($c_i > 99$). The thresholds use
training-set statistics only. Table~\ref{tab:popularity-wise-performance}
reports ICEGR with SQE-SFT across these buckets.


\begin{table}[htbp]
  \caption{Popularity-stratified performance at cutoff 20. Relative gains over Standard SFT
in parentheses.}
  \label{tab:popularity-wise-performance}
  \centering
  \small
  \setlength{\tabcolsep}{2pt}

  \newcommand{\tablesevenmetric}[2]{%
    \begin{tabular}[c]{@{}c@{}}#1\\[-1pt](#2)\end{tabular}%
  }

  \begin{tabular}{@{}
    p{0.12\columnwidth}
    p{0.18\columnwidth}
    p{0.22\columnwidth}
    p{0.22\columnwidth}
    p{0.22\columnwidth}
    @{}}
    \toprule
    \textbf{Bucket}
      & \textbf{Pairs}
      & \textbf{Recall@20}
      & \textbf{MRR@20}
      & \textbf{NDCG@20} \\
    \midrule
    Tail
      & 6,126
      & \tablesevenmetric{0.2088}{+21.22\%}
      & \tablesevenmetric{0.0556}{+38.83\%}
      & \tablesevenmetric{0.0889}{+30.15\%} \\
    Low
      & 4,211
      & \tablesevenmetric{0.4531}{+36.17\%}
      & \tablesevenmetric{0.1449}{+54.84\%}
      & \tablesevenmetric{0.2125}{+46.01\%} \\
    Mid
      & 7,295
      & \tablesevenmetric{0.5934}{+11.31\%}
      & \tablesevenmetric{0.2098}{+17.39\%}
      & \tablesevenmetric{0.2947}{+14.79\%} \\
    Hot
      & 5,486
      & \tablesevenmetric{0.6864}{+4.78\%}
      & \tablesevenmetric{0.2818}{+10.33\%}
      & \tablesevenmetric{0.3727}{+7.89\%} \\
    \bottomrule
  \end{tabular}
\end{table}

\section*{Ethical Considerations}

We use aggregated and de-identified query--item interaction logs
collected and processed under institutional authorization. Before
analysis, we remove raw user identifiers, device identifiers, and
other directly identifying fields. Sensitive free-text queries are
excluded from the research corpus, and access to the resulting data is
restricted to authorized researchers.

Because click-, payment-, and GMV-based signals may amplify
popularity and commercial bias, we constrain preference construction
using query--item relevance and evaluate performance separately for
long-tail queries and low-exposure items. Synthetic queries are used
solely to improve supervision coverage and are filtered for semantic
relevance, attribute consistency, and duplication.

The online evaluation was conducted as a controlled A/B test in
accordance with the applicable approval and monitoring procedures. The
system is intended for e-commerce search and is not designed for
high-impact individual decisions. We monitor the metrics actually
reported in this paper and maintain rollback and incident-response
procedures for deployment.

\bibliographystyle{ACM-Reference-Format}
\bibliography{references}

\end{document}